\documentclass[12pt]{article}
\usepackage{graphicx}
\begin{document}
Simple tools for extrapolations of human mortality in rich countries

\bigskip
Dietrich Stauffer

\bigskip
Institute for Theoretical Physics, Cologne University, D-50923 K\"oln, Euroland

\bigskip
Abstract: Suitable assumptions for the Gompertz mortality law take into account
the break in the time development observed recently by  Wilmoth et al. They
show how a drastic reduction in the birth rate and improved living conditions
lead to a drastic increase in the fraction of old people in the population, 
and how immigration of half a percent of the population per year can mostly
stop this increase.

\bigskip
Human life expectancies increased and birth rates decreased, in the 20th century
in most rich countries, and this may happen soon elsewhere [1]. Predictions
for the future [2-4] are usually extrapolations of the trends of the past and
thus are problematic if extended over centuries as in ref.4. Recent analysis
of the age of the oldest people in a finite population versus time [5], of 
survival probabilities as a function of life expectancy at birth [6], or of
survival probabilities as a function of survival probability up to age 40 [7]
raise the possibility of sharp breaks in these curves. By definition, no 
extrapolation beyond the next (unknown) break is possible, if these breaks are
infinitely sharp (if infinitely accurate data would be available). 
Now we try to describe these effects by a simple model with a limited number
of parameters, and show the versatility of this model by an extrapolation
of the fraction of old people under various hypotheses. While some aspects are 
motivated by German data, the one-page computer program could easily be adapted 
to other rich countries with birth rates below replacement level and is
available from stauffer@thp.uni-koeln.de. 

Human adult mortalities between the ages $x$ of 30 and 90 years follow roughly 
an exponential increase $\propto \exp(bx)$, known as the Gompertz law since the
19th century. Much of the trends of the past centuries in various 
countries were summarized by the universality law [8]:
$$ \mu/b = A \, e^{b(x-X)}$$
where $\mu$ is the mortality function,
$A$ and $X \simeq 103$ years are universal over the whole human
species, and only the slope $b$ increases with increasing progress. The time
unit is one year throughout.
(Also Thatcher [9] noted that human mortality at about 100 years is roughly
constant.) We take $b$ to increase linearly with time from 0.07 in 1821 to
0.093 in 1971, and from then on to remain constant. Fig.1a compares the 
mortality function of German women with this simple assumption. We see that
our analysis is valid only at older ages and only for 
the last decades of the 20th century in rich countries when child mortality 
was already low. Fig.1b compares the life expectancy at age 65
for German men with simulations of the model explained below.

\begin{figure}[hbt]
\begin{center}
\includegraphics[angle=-90,scale=0.34]{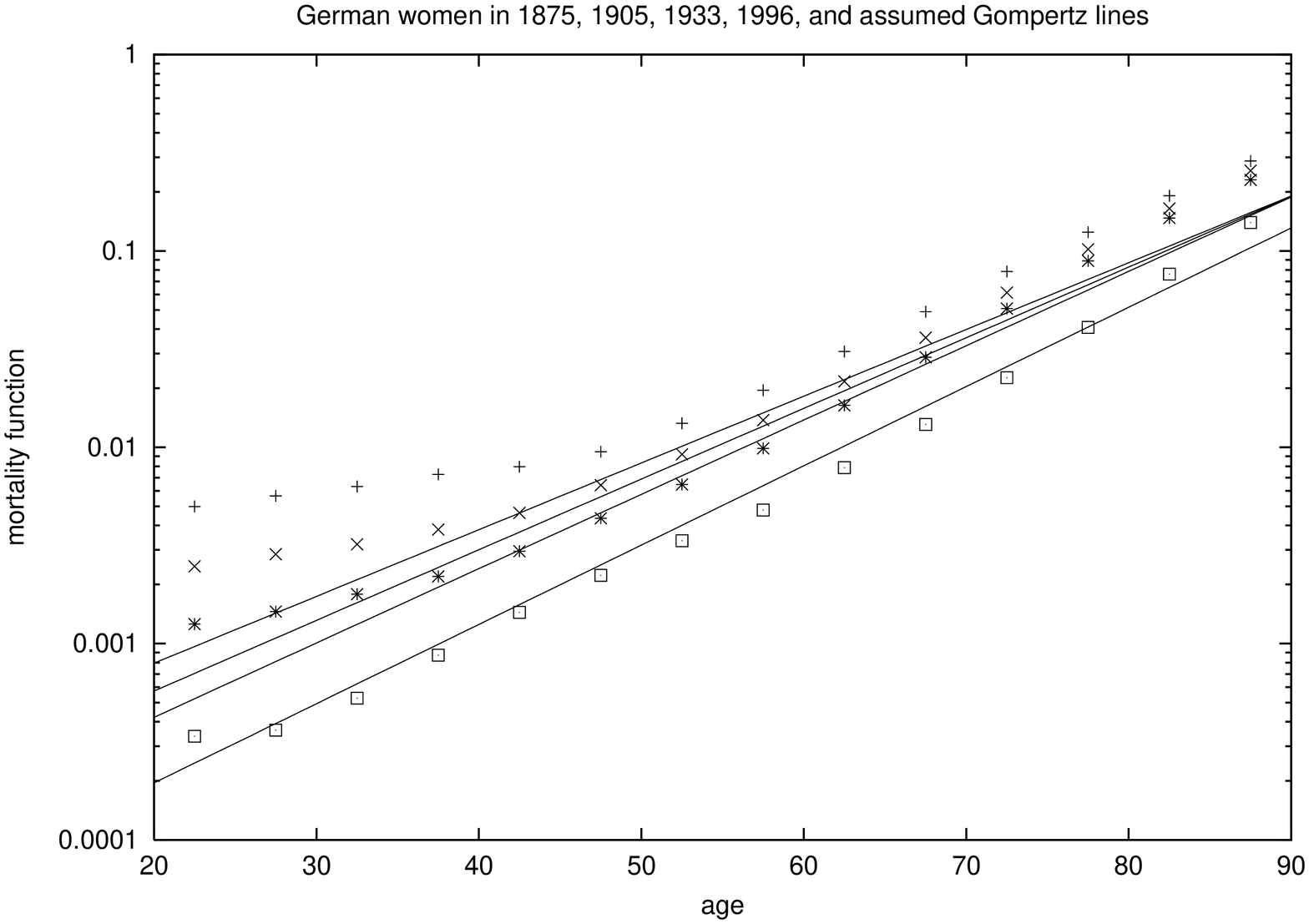}
\includegraphics[angle=-90,scale=0.34]{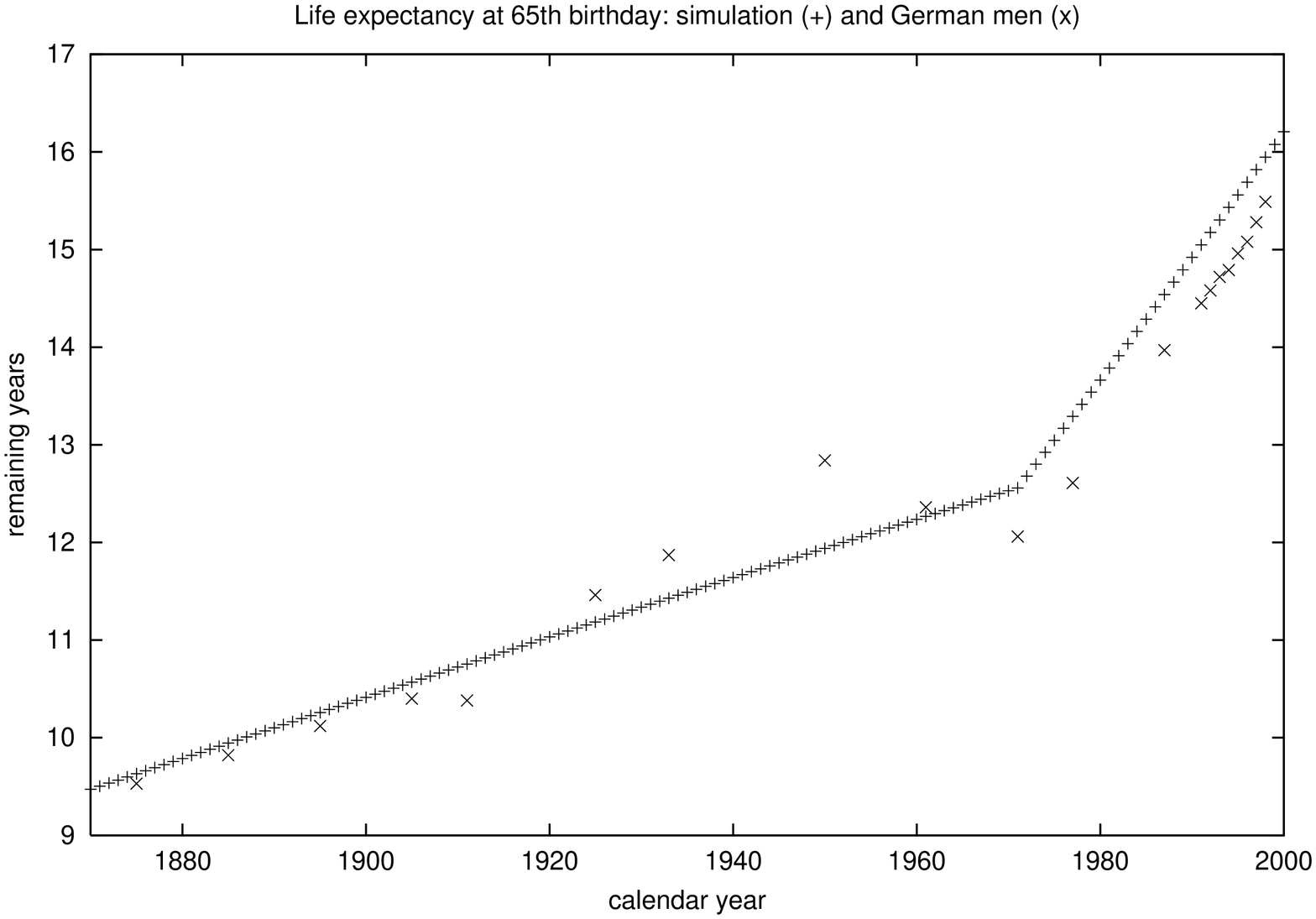}
\end{center}
\caption{
a) German mortality functions and Gompertz law, for 1875, 1905, 1933, and 
1996. b) Remaining life expectancy for German men at age 65, and simulation
results from the simple model presented later.
}
\end{figure}

(In the above equation, the mortality function $\mu = -d \ln S(x)/dx$ is also
called the hazard factor or force of mortality [10] and is approximated by
$\ln[S(x-1/2)/S(x+1/2)]$ where $S(x)$ (often denoted as $l(x)$) is the 
probability to survive from birth to age $x$. 
For centenarians, downward deviations of $\mu$ from the Gompertz law have been 
reported [10] but an extrapolation to zero systematic errors (from age
overstating) seems lacking.)

\begin{figure}[hbt]
\begin{center}
\includegraphics[angle=-90,scale=0.5]{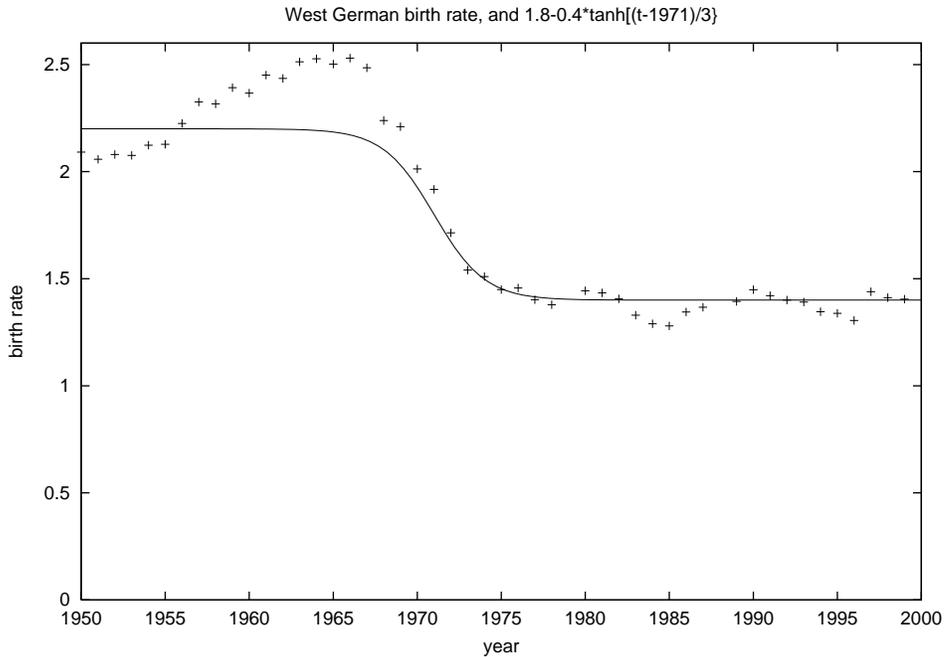}
\end{center}
\caption{
West German birth rate in recent decades, and fitted tanh formula.
}
\end{figure}

However, these assumptions are still wrong since they lead for increasing $b$
(increasing progress) to a {\it decrease} of the average maximum age in a finite
population. (This maximum age $x_{max}$ is defined as the age at which the
reciprocal survival probability $1/S(x_{max})$ reaches the population size [11];
we ignore the growth of the population with time and also the variation of the 
mortality parameters during the lifetime of an individual.) In reality, 
$x_{max}$ increased
with increasing progress during the last century [5], page 46 in [1]. 

To allow for $x_{max}$ to increase with calendar year $t$, we allow the
characteristic age $X$ to increase with time after 1971, by generalizing 
$X = 103$ to 
$X = 103 + 0.15(t-1971)$. This generalization is already included in Fig.1 and
slightly separates the lowest straight line from the three higher (earlier)
ones. Now $x_{max}$ increases with increasing $t$, similar to reality (page 46
in [1].) 

\begin{figure}[hbt]
\begin{center}
\includegraphics[angle=-90,scale=0.34]{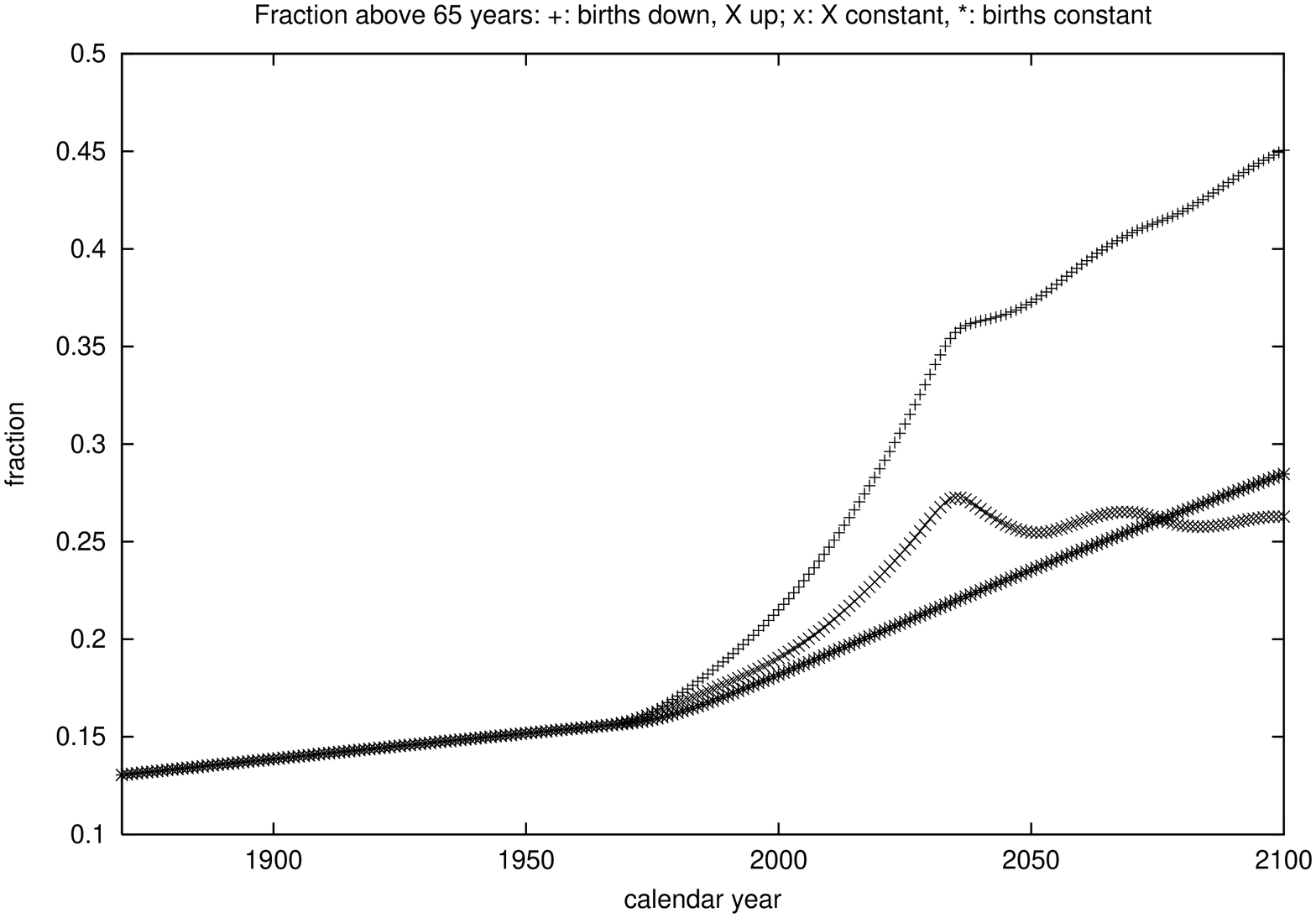}
\includegraphics[angle=-90,scale=0.34]{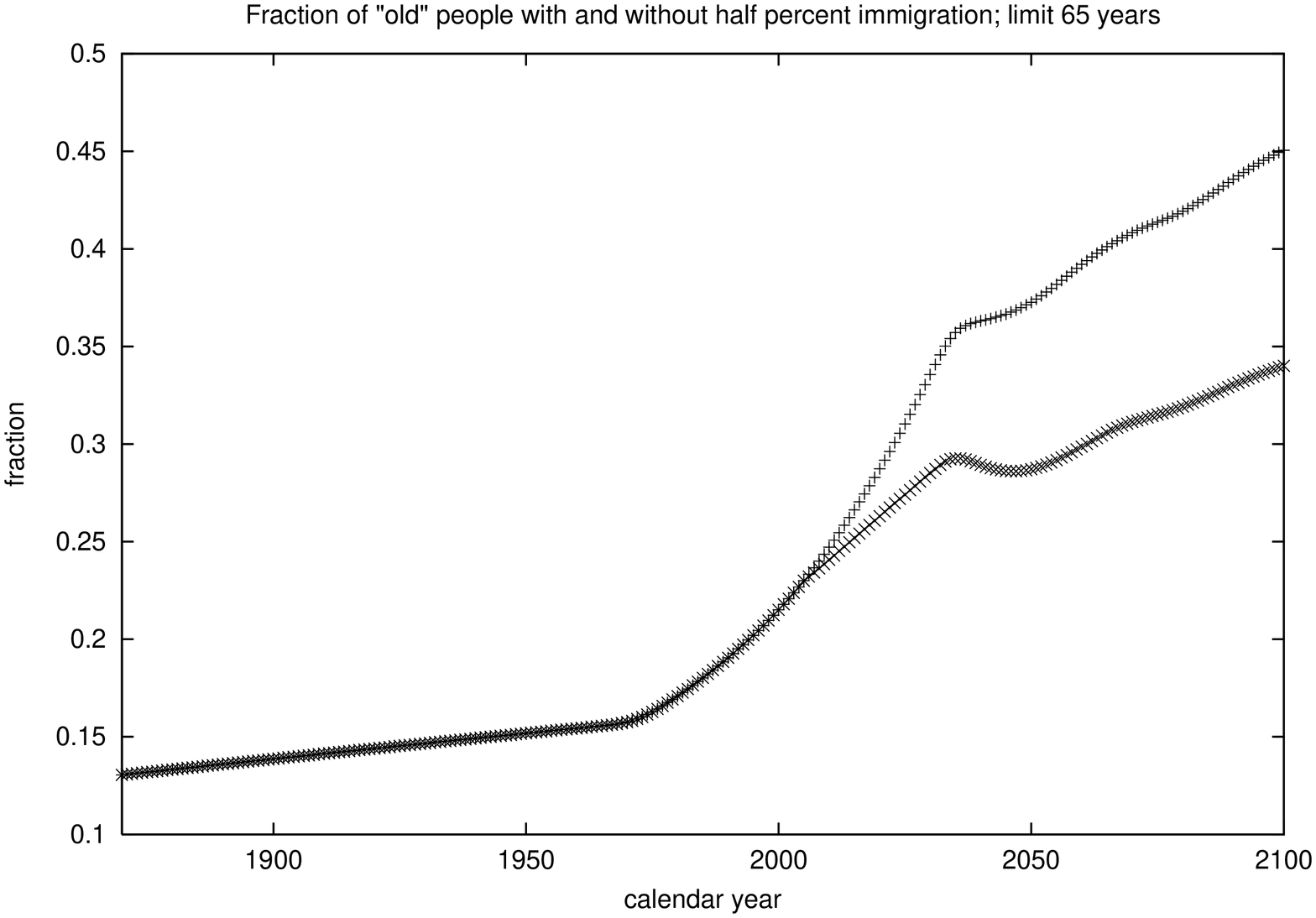}
\end{center}
\caption{
a) Fraction of old people in the computer simulation (x), together with the
hypothetical development had the birth rate not dropped (+) or had the
characteristic age $X$ not increased (*). b) Same higher curve from part a,
compared with lower curve assuming after 2005 immigration to stabilize 
population.
}
\end{figure}

The two effects, the increase of $b$ before the year 1971, and the increase of
$x$ thereafter, describe the two empirical observations reviewed recently by
Yashin et al [12]: During the first half the the 20th century the survival
curves $S(x)$ became more rectangular (``compression of mortality''), but
later this rectangularization stopped and is replaced by a shifting of the
old-age survival curve towards older ages at constant shape.
The interquartile death rate is the number of years between $S(x)=0.75$ and 
$S(x)=0.25$. In my calculations it first declines and then stays constant
similar to Swedish reality 
(Fig.4a of Wilmoth and Horiuchi [5], if deaths before age 30 are ignored).

\bigskip
For the birth rate we assume that pregnancies happen at ages 20 to 40 only,
with equal probability in each of these 20 years. We first adjust a birth 
rate $m=2.17$ such that for $b = 0.07$ (a realistic value
for earlier centuries [9]) a stationary population is reached.
(Note that ``births'' here count only babies reaching adult age; child mortality
was ignored here from the beginning.) The birth rate  then increased linearly
in time to 2.2 for the year 1971. Then very rapidly it decreases to 1.4 and 
stays there, according to a hyperbolic tangent formula 
$$ m(t) = 1.8 - 0.4 \cdot \tanh[(t-1971)/3] $$
compared in Fig.2 with West German reality.

Thus, compatible with Wilmoth et al [5], the year 1971 is taken as both the
year when rectangularization was replaced by shifting in $S(x)$, and then
the birth rate had its strongest decline. These sharp changes in the higher 
time derivatives may be missing in traditional demographic extrapolations [2-4]
but could be easily implemented in a Fortran program of about 50 lines. Thus
these methods could be used in more realistic fits and extrapolations for 
specific countries. 

Figure 3 illustrates from these assumptions the future development for  the 
fraction of
``old'' people, the limit taken at 65. About this ``age quake'' 
much has been talked, not done. In Germany, for example, three quarters of the
population prefer to retire before the age of 60, and a new immigration law is 
not even called by that name. Figure 3a also shows what would have happened if
the birth rate would not have dropped, or if the characteristic age $X$ would 
not increase. Then, the increase in the fraction of old people would be
much weaker. More realistically, Fig.3b shows the changes made by an immigration
of people aged 6 to 40, amounting to half a percent of the total population
per year and roughly stabilizing the population. 

In summary, the possible breaks in the time development of demographic 
parameters were included in a simple approximation which could be incorporated
into more detailed traditional models of human population dynamics. Simulations 
are in preparation [13] to check whether these phenomenological assumptions can 
be reproduced by more microscopic models [14].

I thank S. Moss de Oliveira for discussions.
 
\parindent 0pt
\bigskip
[1] Wachter, K.W., Finch, C.E., 1997. {\it Between Zeus and the Salmon. The 
Biodemography of Longevity}, National Academy Press, Washington DC; see  
also La Recherche 322 (various authors), July/August 1999;
Nature 408, No. 680 (various authors), November 9, 2000.

[2] Bomsdorf, E., 1993. {\it Generationensterbetafeln f\"ur die 
Geburtsjahrg\"ange 1923-1993: Modellrechnungen f\"ur die Bundesrepublik 
Deutschland}, Verlag Josef Eul, K\"oln.

[3]  Tuljapurkar, S., Li, N., Boe, C., 2000. A universal pattern of mortality 
decline in the G7 countries. Nature 405, 789-792.

[4] Olshansky, S.J., Carnes B.A., D\'esesquelles A., 2001.
Demography - Prospects for human longevity Science 291, 1491-1492.

[5] Wilmoth, J.R., Deegan, L.J., Lundstr\"om, H.,  Horiuchi, S., 2000. 
Increase of maximum life-span in Sweden, 1861-1999.  Science 289, 2366-2368; 
Wilmoth, J.R., Horiuchi, S., 1999. Rectangularization revisited: Variability of 
age at death within human populations. Demography 36, 475-495;
Wilmoth, J.R., Lundstr\"om, H., 1996. Extreme longevity in five countries - 
Presentation of trends with special attention to issues of data quality.
Eur. J. Population 12, 63-93.

[6] Azbel, M. Ya., 1999. Phenomenological theory of mortality evolution: Its 
singularities, universality, and superuniversality. PNAS USA 96, 3303-3307.

[7] Azbel, M.Ya., 1999. Empirical laws of survival and evolution: Their 
universality and implications. PNAS USA 96, 15368-15373; 2001. Phenomenological 
theory of survival.  Physica A  297, 235-241.

[8] Strehler, B. L., Mildvan, A.S. 1960. General theory of mortality
and aging, Science 132, 14-21;
L.A. Gavrilov, L.A., Gavrilova, N.S. 1991. {\it The Biology of Life Span},
Harwood Academic Publisher, Chur and 2001. The reliability theory of aging and
longevity, J. Theor. Biology 213, 527-545;
Azbel, M. Ya., 1996. Unitary mortality law and species-specific age.
Proc. Roy. Soc. B 263, 1449-1454.

[9] Thatcher, A.R., 1999: The long-term pattern of adult mortality and the 
highest attained age. J. Roy. Statist. Soc. A 162, 5-30,  and priv.comm.

[10] Thatcher, A.R., Kannisto, V., Vaupel, J.W., 1998. {\it The Force 
of Mortality at Ages 80 to 120}: Odense University Press, Odense.
Robine, J.-M., Vaupel, J.W., 2001. Supercentenarians: slower ageing
individuals or senile elderly?,  Exp. Gerontology 36, 915-930 (2001)

[11] Finch C.E., Pike, M.C., 1996. Maximum life span predictions from the 
Gompertz mortality model. J. Gerontology (Biol. Sci.) 51 A, B 183-194.

[12] Yashin, A.I., Begun, A.S., Boiko, S.I., Ukraintseva, S.V., Oeppen, J., 
2001. The new trends in survival improvement require a revision of traditional
gerontological concepts., Exp. Gerontology 37, 157-167.

[13] Cebrat, S., 2002, private communication

[14] de Oliveira, P.M.C., Moss de Oliveira, S., Stauffer, D., Cebrat, S. 1999.
Penna ageing model and improvement of medical care in 20th century.
Physica A 273, 145-149; Niewczas, E., Cebrat, S., Stauffer, D. 2000.
The Influence of the Medical Care on the Human Life Expectancy in
20th Century and the Penna Ageing Model. Theory Biosci. 119, 122-131; 
Moss de Oliveira, M., de Oliveira, P.M.C., Stauffer, D., 1999. 
{\it Evolution, Money, War and Computers}, Teubner, Stuttgart and Leipzig;
Stauffer D., Radomski, J.R., 2001. Social effects in simple computer model of
aging, Exp. Gerontology 37, 175-180.
\end{document}